\documentclass[lettersize,journal]{IEEEtran}
\usepackage{booktabs}
\usepackage{longtable}
\usepackage[flushleft]{threeparttable}
\usepackage[flushleft]{threeparttablex}
\usepackage{multirow}
\usepackage{multicol}
\usepackage{amsmath,amssymb,amsfonts}
\usepackage{algorithmic}
\usepackage{algorithm}
\usepackage{array}
\usepackage{textcomp}
\usepackage{stfloats}
\usepackage{url}
\usepackage{verbatim}
\usepackage{graphicx}
\usepackage{cite}
\usepackage{eurosym}
\usepackage{acronym}
\usepackage[mathscr]{eucal}
\usepackage{xcolor}
\usepackage{subfigure}

\usepackage{accents}
\newcommand{\ubar}[1]{\underaccent{\bar}{#1}}

\newcommand{\DA}{\rm{DA}}
\newcommand{\SR}{\rm{SR}}

\newcommand{\mT}{\mathscr{T}}

\newcommand{\mP}{\mathscr{P}}
\newcommand{\mR}{\mathscr{R}}
\newcommand{\mD}{\mathscr{D}}

\usepackage{nomencl}
\newcommand{\nomunit}[1]{%
\renewcommand{\nomentryend}{\hspace*{\fill}#1}}
\makenomenclature

\usepackage{etoolbox}
\renewcommand\nomgroup[1]{%
  \item[\bfseries
  \ifstrequal{#1}{I}{Indexes and Sets}{%
  \ifstrequal{#1}{P}{Parameters}{%
  \ifstrequal{#1}{V}{Continuous Variables}{%
  \ifstrequal{#1}{Z}{Binary Variables}{}}}}%
]}

\graphicspath{{figs/}} 

\acrodef{dam}[DAM]{Day Ahead Market}
\acrodef{srm}[SRM]{Secondary Reserve Market}
\acrodef{asm}[ASM]{Ancillary Service Market}
\acrodef{sr}[SR]{Secondary Reserve}
\acrodef{rto}[RTO]{Real-Time Operation}
\acrodef{rt}[RT]{Real-Time}
\acrodef{ess}[ESS]{Energy Storage System}
\acrodef{bess}[BESS]{Battery Energy Storage System}
\acrodef{res}[RES]{Renewable Energy Source}
\acrodef{ndrs}[ND-RES]{Non-dispatchable Renewable Energy Sources}
\acrodef{stu}[STU]{Solar Thermal Unit}
\acrodef{vpp}[VPP]{Virtual Power Plant}
\acrodef{wpp}[WPP]{Wind Power Plant}
\acrodef{pv}[PV]{Photo-Voltaic}
\acrodef{opf}[OPF]{Optimal Power Flow}
\acrodef{pf}[PF]{Power Flow}
\acrodef{milp}[MILP]{Mixed Integer Linear Programming}
\acrodef{minlp}[MINLP]{Mixed Integer non-Linear Programming}
\acrodef{tso}[TSO]{Transmission System Operator}
\acrodef{so}[SO]{System Operator}
\acrodef{pcc}[PCC]{Point of Common Coupling}
\acrodef{ree}[REE]{Red El{\'e}ctrica de España}
\acrodef{afrr}[aFRR]{Automatic Frequency Restoration Reserve}
\acrodef{fcr}[FCR]{Frequency Containment Reserve}
\acrodef{rvpp}[RVPP]{RES-only Virtual Power Plant}
\acrodef{gams}[GAMS]{General Algebraic Modeling System}
\acrodef{gdx}[.gdx]{GAMS Data eXchange}
\acrodef{picasso}[PICASSO]{Platform for the International Coordination of Automated Frequency Restoration and Stable System Operation}
\acrodef{so}[SO]{Stochastic Optimization}
\acrodef{ro}[RO]{Robust Optimization}
\acrodef{aro}[ARO]{Adaptive Robust Optimization}
\acrodef{saro}[SARO]{Stochastic ARO}

\acrodef{ccg}[CCG]{Column Constraint Generation}
\acrodef{rtm}[RTM]{Real-time Market}
\acrodef{ngm}[NGM]{Natural Gas Market}

\begin{document}

\title{Single-level Robust Bidding of Renewable-only Virtual Power Plant in Energy and Ancillary Service Markets for Worst-case Profit Optimization}

\author{Hadi Nemati,~\IEEEmembership{Student,~IEEE}, Pedro S{\'a}nchez-Mart{\'i}n, Ana Baringo, {\'A}lvaro Ortega,~\IEEEmembership{Member,~IEEE \vspace{-8mm}} 
\thanks{This project has received funding from the European Union’s Horizon 2020 research and innovation programme under grant agreement No 883985}
}



\maketitle

\begin{abstract}
This paper proposes a novel single-level robust mathematical approach to model the \ac{rvpp} bidding problem in the simultaneous \ac{dam} and \ac{srm}. The worst-case profit of \ac{rvpp} due to uncertainties related to electricity prices, \ac{ndrs} production, and flexible demand is captured. In order to find the worst-case profit in a single-level model, the relationship between price and energy uncertainties leads to some non-linear constraints, which are appropriately linearized. The simulation results show the superiority of the proposed robust model compared to those in the literature, as well as its computational efficiency.

\end{abstract}

\vspace{-1em}

\section{Introduction}

\vspace{-2mm}
\subsection{Motivation}
\IEEEPARstart{T}{he} 
penetration of \acp{ndrs} has experienced a remarkable growth in the last decades. However, the stochastic nature of these sources implies that \acp{ndrs} are less reliable when it comes to predictable and controllable power injection over a given period of time~\cite{gulotta2023short}. This makes \acp{ndrs} participation in the energy and \ac{asm} difficult, as failure to meet with the contracted energy and reserve in the market will lead to penalties if not suspension from future market activities. However, by integrating multiple portfolios of \acp{ndrs} and other flexible assets as an \ac{rvpp}, the performance and competitiveness of \acp{ndrs} in these markets can be significantly improved~\cite{yang2023optimal}.

The viability of \ac{rvpp} depends on its economic performance, related to benefits and costs. Different markets bring different benefits according to the bidding/offering ability of \ac{rvpp} and its ability to provide what is promised~\cite{alvaroPOSYFT}. However, in addition to the internal uncertainties of \ac{rvpp} units in their production and demand, there are various external uncertainties in the markets, such as the energy and reserve electricity price uncertainties~\cite{zhang2021robust}. Therefore, the development of bidding approaches for \ac{rvpp} participation in different markets taking into account the characteristics of \ac{rvpp} units, market rules, and internal and external uncertainties has at most important for \ac{rvpp} operators and researchers~\cite{venegas2022review}.

\vspace{-4mm}
\subsection{Literature Review}
Many papers in the literature use mathematical optimization models to capture different uncertainties associated with \ac{vpp} due to ease of implementation, convergence to the global optimum, and computational efficiency of these models~\cite{gao2024review}. In this context, \ac{ro} programming is an efficient way to deal with different sets of uncertainties that vary in their possible values. The goal of \ac{ro} is to find the worst case of the optimization problem to minimize the negative impact of uncertainties on the solution~\cite{naval2021virtual}. However, the definition of the worst case can vary depending on how the optimization is implemented, whether it is single-level or multi-level, and can lead to different solutions in each approach. The authors in~\cite{ZHANG2023108558, bafrani2022robust, ju2019cvar, rahimiyan2015strategic, wang2023optimal, oskouei2021strategic, silva2024light, Hadi} develop a single-level optimization problem for the \ac{vpp} market bidding problem to find the worst case of energy of \acp{ndrs}.
{The literature addresses \ac{vpp} scheduling and bidding problems by considering different uncertainty characterizations in demand~\cite{ZHANG2023108558, bafrani2022robust, ju2019cvar, oskouei2021strategic, Hadi}, \ac{ndrs} production~\cite{bafrani2022robust, ju2019cvar, rahimiyan2015strategic,wang2023optimal, oskouei2021strategic, silva2024light, Hadi}, and electricity price~\cite{rahimiyan2015strategic,wang2023optimal, oskouei2021strategic, Hadi}, focusing on multi-market~\cite{ZHANG2023108558, bafrani2022robust, rahimiyan2015strategic, wang2023optimal, oskouei2021strategic, silva2024light, Hadi}, multi-objective~\cite{ju2019cvar}, and multi-energy models~\cite{wang2023optimal, oskouei2021strategic}.} The main advantages of the mentioned single-level \ac{ro} programming in~\cite{ZHANG2023108558, bafrani2022robust, ju2019cvar, rahimiyan2015strategic, wang2023optimal, oskouei2021strategic, silva2024light, Hadi} are the possibility to consider multiple uncertainties, simplicity of implementation, global optimality, and calculation efficiency. However, a simplified definition of the worst case of energy for the severe scenarios is implemented. In fact, the worst case of energy defined for \acp{ndrs} in the above papers does not lead to the worst condition of profit, considering the possibility of different values of electricity prices. For instance, in a case where the electricity price is low in a certain period, even though the energy of a \ac{ndrs} can deviate significantly in this period, the resulting loss for \ac{rvpp} might not be significant compared to a period with much higher electricity price and average or low energy deviation.

Multi-level \ac{ro} models provide more flexibility to find the actual worst-case of the \ac{vpp} bidding problem compared to single-level models. This is due to the definition of a new level for the optimization problem that models the behavior of uncertain parameters (both electrcity price and energy uncertainties). Therefore, the objective function of this level can be defined to find the worst case of energy or profit of \ac{vpp}. In addition, another level for the problem can be included to define the corrective or remedial actions after the occurrence of uncertainties.

{The literature on multi-level models proposes mathematical techniques, including Adaptive~\cite{khojasteh2023novel, li2022robust} and Stochastic~\cite{zhang2021robust, baringo2018day, kong2020robust} \ac{ro} to account for various uncertainties in \ac{ndrs} production~\cite{zhang2021robust, baringo2018day, khojasteh2023novel, kong2020robust, li2022robust}, demand~\cite{khojasteh2023novel, kong2020robust}, electricity prices~\cite{zhang2021robust, baringo2018day}, and reserve deployment or dispatch order of \ac{tso}~\cite{zhang2021robust, baringo2018day, li2022robust}. The proposed models are implemented for the multi-market participation of \ac{vpp}~\cite{zhang2021robust, baringo2018day, khojasteh2023novel} and for multi-energy \acp{vpp}~\cite{kong2020robust}. Different techniques, including Benders and other decomposition techniques~\cite{zhang2021robust, khojasteh2023novel}, the \acl{ccg} algorithm~\cite{baringo2018day}, and improved versions of these algorithms~\cite{kong2020robust, li2022robust}, are proposed to accelerate the solution time of the optimization problem.} The main limitations of the multi-level approaches in general, and in the above works in particular, are the complexity of programming and the fact that the size of the problem grows with the number of iterations in the solving procedure. {In addition, they usually imply long computational times, which can compromise applications such as sensitivity analysis.}

\vspace{-3mm}
\subsection{Approach and Contributions}
To avoid the difficulties of implementing a multi-level model and the computational complexity for practical applications, this paper models the worst-case profit of \ac{rvpp} against uncertainties by means of a novel single-level \ac{milp} problem. The equations related to the uncertain parameters in the objective function of the optimization problem (equations related to the energy and reserve price uncertainties) as well as the equations related to the uncertain parameters in the constraints (equations related to the \ac{ndrs} and demand uncertainties) are defined by developing the approach in~\cite{bertsimas2004price, rahimiyan2015strategic, Hadi} and by developing on the idea from the big M method~\cite{floudas1995nonlinear}. The proposed implementation of robust constraints makes it possible to capture the relationship between different uncertain parameters in the objective and constraints of the optimization problem, and to find the exact worst case profit of \ac{rvpp}. Finally, defining the relationship between uncertain parameters in order to find the worst case leads to some non-linear constraints, which are linearized by using well-established methods. 
 
The contributions of this paper are threefold:

\begin{itemize}
   \item {Modeling the worst-case profit robustness of an \ac{rvpp} with a single-level \ac{minlp} model}. As opposed to other single-level models in the literature, the proposed model maximizes the expected profit of \ac{rvpp} for the simultaneous \ac{dam} and \ac{srm} participation against the worst-case profit robustness of different uncertainties on prices and energy (\ac{ndrs} production and demand). 
    
    \item {Addressing the non-linear couplings between various uncertainties} within the optimization problem, and subsequently formulating an equivalent \ac{milp} problem for the initial \ac{minlp} one. 
    
    \item The proposed single-level \ac{milp} model has high computational efficiency and simpler implementation compared to the multi-level optimization models in the literature.
\end{itemize}

\vspace{-3mm}
\subsection{Paper Organization}
The reminder of the paper is organized as follows. A conceptual comparison of energy and profit robustness approaches is presented in Section~\ref{sec:Energy vs. profit}. The proposed single-level robust bidding problem of \ac{rvpp} for \ac{dam} and \ac{srm} participation is formulated in Section~\ref{sec:Formulation}. An illustrative example is given in Section~\ref{sec: Example} to show the performance of the proposed robust model in finding the worst-case profit. The simulation results are presented in Section~\ref{sec:Simulation}. Finally, the conclusions are drawn in Section~\ref{sec:Conclusion}.

\vspace{-2mm}
\section{Comparing Energy and Profit Robustness}
\label{sec:Energy vs. profit}
\begin{figure}[hbt!]
    \centering  \includegraphics[width=\columnwidth]{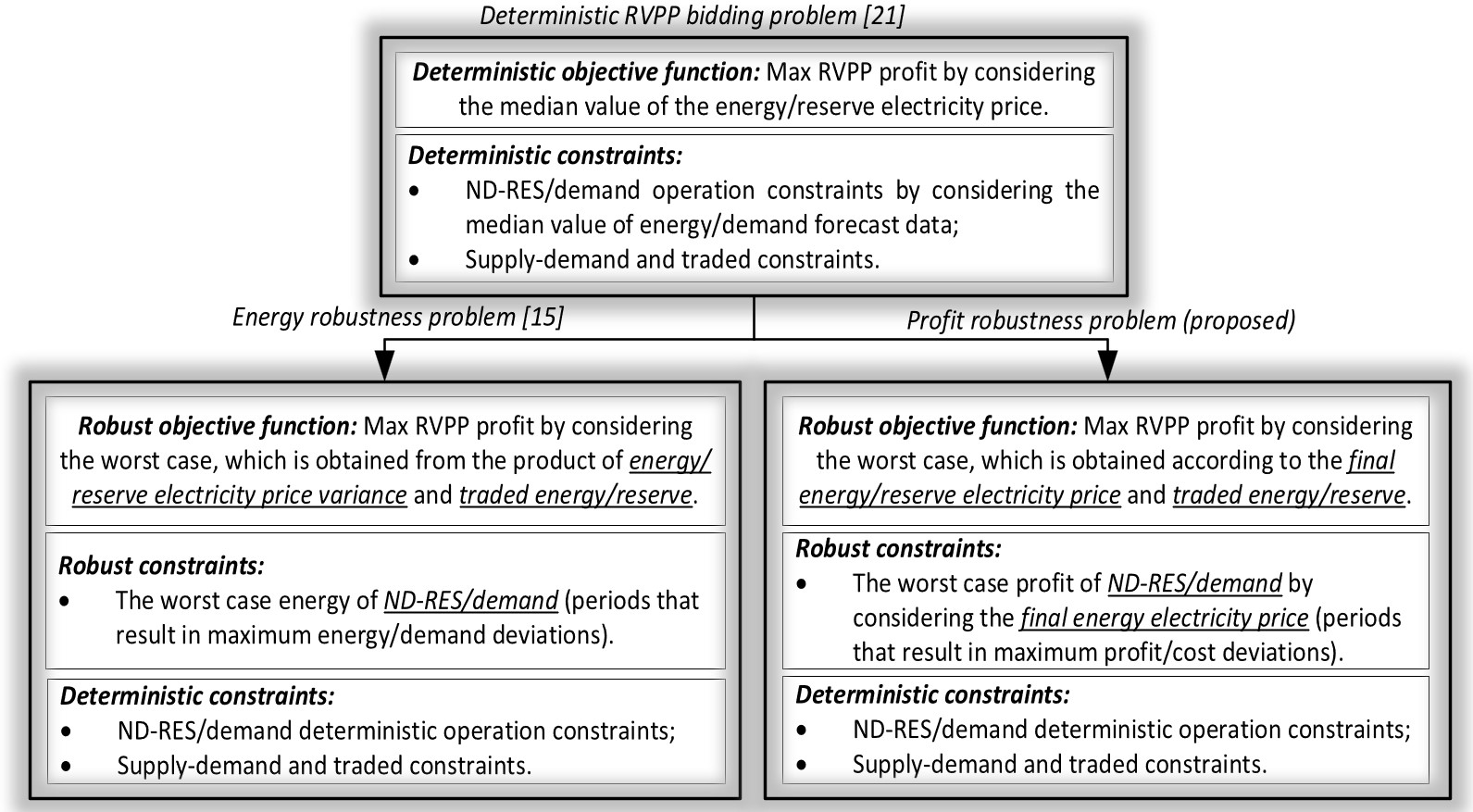}
    \vspace{-2em}
    \caption{A comparison between the energy and profit robustness approaches.}
    \label{fig:EnergyvsProfit_Robustness_fig}
\end{figure}

Figure~\ref{fig:EnergyvsProfit_Robustness_fig} shows the structure of a deterministic \ac{rvpp} bidding problem and a comparison between the energy and the profit robustness approaches. In the deterministic approach, a single value (usually the median or average) of the forecast data is considered to solve the optimization problem. The constraints are related mainly to the operation of the \ac{rvpp} units, and supply-demand balance~\cite{oladimeji2022optimal}. When considering the uncertainties, depending on whether the uncertainties affect the objective function or the constraints of the optimization problem, different sets of constraints need to be defined in each of the \ac{ro} approaches. The uncertainties related to the energy/reserve electricity price affect the objective function of the optimization problem, whereas the uncertainties associated with the \ac{ndrs} generation and demand consumption affect the constraints.

\vspace{-2mm}

{In the energy robustness approach, those periods that result in more deviation of the energy/reserve electricity price variance multiplied by the total traded energy/reserve of \ac{rvpp} are selected as the worst-case scenarios of the electricity price~\cite{Hadi}. 
In the energy robustness constraints, the periods that have higher deviation of energy are selected as the worst case of \acp{ndrs} production or demand regardless of the electricity price.}

In the profit robustness approach proposed in this paper and for the uncertain parameters in the objective function of the optimization problem (energy/reserve electricity price), the worst case is defined according to the final value of the energy/reserve electricity price by means of binary variables. The final value of the energy electricity price is also used to calculate the worst case of profit/cost of each unit (uncertainty of \ac{ndrs} and demand in the constraints of the optimization problem). For this purpose, the final energy electricity price is multiplied by the energy variable of \ac{ndrs}/demand and is limited by the profit reduction effect due to \ac{ndrs}/demand uncertainty. 

In the following section, the proposed profit robustness approach is formulated as a single-level optimization problem. In Section~\ref{sec: Example}, these two approaches are compared using an illustrative example.

\section{Profit Robustness Formulation}
\label{sec:Formulation}

\subsection{Nomenclature}
This subsection presents the notation and nomenclature used in the remainder of the paper.

\vspace{2mm}
\noindent \textbf{General Notation Concepts}
\begin{itemize}
    \item An uncertain parameter with a tilde symbol denotes the median value in the forecast distribution, representing a point where half of the observations are lower ($\tilde{A}$);
    \item the hat/inverse hat symbol on uncertain parameters signifies the greatest positive/negative permitted deviation from the forecast's median ($\hat{A}$, $\check{A}$);
    \item parameters with an upper/lower bar represent their upper/lower bounds of parameter $A$ ($\bar{A}$, $\ubar{A}$);
    \item upward/downward arrows indicate up/down direction of regulation in variables and parameters ($a^{\uparrow}$, $A^{\uparrow}$/$a^{\downarrow}$, $A^{\downarrow}$).
\end{itemize}
\vspace{-7mm}
\setlength{\nomitemsep}{0.05cm}

\nomenclature[I, 01]{$d \in \mathscr{D}$}{Set of demands \nomunit{}}
\nomenclature[I, 01]{$p \in \mathscr{P}$}{Set of daily load profiles \nomunit{}}
\nomenclature[I, 01]{$r \in \mathscr{R}$}{Set of \acp{ndrs} \nomunit{}}
\nomenclature[I, 01]{$t \in \mathscr{T}$}{Set of time periods \nomunit{}}
\nomenclature[I, 03]{$\Xi^{\rm{DA+SR}}$}{Set of decision variables of \ac{dam} and \ac{srm} \nomunit{} \vspace{5pt}}

\nomenclature[P, 01]{$C_{d,p} \;$}{Cost of load profile $p$ of demand $d$ \nomunit{[€]}}  
\nomenclature[P, 01]{$C_r^R \;$}{Operation and maintenance costs of \ac{ndrs} $r$ \nomunit{[€/MWh]}}
\nomenclature[P, 01]{$E_d \;$}{Energy consumption of demand $d$ throughout the planning horizon \nomunit{[MWh]}}
\nomenclature[P, 01]{$M \;$}{Very big positive value \nomunit{[€]}}
\nomenclature[P, 01]{$P_d$}{Power consumption of demand $d$ \nomunit{[MW]}}
\nomenclature[P, 01]{$P_{d,p,t}$}{Profile $p$ of demand $d$ prediction during period $t$ \nomunit{[MW]}}
\nomenclature[P, 01]{$P_r$}{Power production of \ac{ndrs} $r$ \nomunit{[MW]}}
\nomenclature[P, 01]{$P_{r,t}$}{\ac{ndrs} $r$ production prediction during period $t$ \nomunit{[MW]}}
\nomenclature[P, 01]{$R_d$}{Ramp rate of demand $d$ \nomunit{[MW/hour]}}
\nomenclature[P, 01]{$R_{r(d)}^{\rm SR}$}{\ac{sr} ramp rate of \ac{ndrs} $r$ (demand $d$) \nomunit{[MW/min]}}
\nomenclature[P, 01]{$T^{\rm SR} \;$}{Required time for \ac{sr} action \nomunit{[min]}}
\nomenclature[P, 02]{$\beta_{d,t}$}{Percentage of flexibility of demand $d$ during period $t$ \nomunit{[\%]}}
\nomenclature[P, 02]{$ \Gamma^{\rm {DA/SR}}$}{\ac{dam}/\ac{srm} price uncertainty budget \nomunit{[-]}}
\nomenclature[P, 02]{$ \Gamma_{r(d)}$}{\ac{ndrs} $r$ production (demand $d$) uncertainty budget \nomunit{[-]}}
\nomenclature[P, 02]{$ \Delta{t} \;$}{Duration of periods \nomunit{[hour]}}
\nomenclature[P, 02]{$ \varepsilon \;$}{Very small positive value \nomunit{[€]}}
\nomenclature[P, 02]{$ \kappa \;$}{User-defined parameter to set the limit of up reserve traded in the \ac{srm} as a percentage of total power capacity of \ac{rvpp} \nomunit{[\%]}}
\nomenclature[P, 02]{$\lambda_t^{\rm{DA/SR}}$}{\ac{dam}/\ac{srm} price prediction during period $t$ \nomunit{[€/MWh]/[€/MW]}}
\nomenclature[P, 02]{$ \varrho_t \;$}{Coefficient to calculate the ratio of down-to-up reserve requested by the \ac{tso} during period $t$ \nomunit{[\%]}}\vspace{5pt}

\nomenclature[V, 01]{$p_{r(d),t}^{\rm DA}$}{Production of \ac{ndrs} $r$ (consumption of demand $d$) in the \ac{dam} during period $t$ \nomunit{[MW]}}
\nomenclature[V, 01]{$p_t^{\rm DA}$}{Total traded power by \ac{rvpp} in the \ac{dam} during period $t$ \nomunit{[MW]}}
\nomenclature[V, 01]{$r_t^{\rm SR}$}{Total \ac{sr} traded by \ac{rvpp} for different \ac{tso} calls on conditions during period $t$ \nomunit{[MW]}}
\nomenclature[V, 01]{$r_{r(d),t}^{\rm SR}$}{\ac{sr} provided by \ac{ndrs} $r$ (demand $d$) for different \ac{tso} calls on conditions during period $t$ \nomunit{[MW]}}
\nomenclature[V, 01]{$y_t^{(\prime)\rm DA}$}{\ac{rvpp} profit affected by \ac{dam} negative (positive) price uncertainty during period $t$ \nomunit{[€]}}
\nomenclature[V, 01]{$y_t^{\rm SR}$}{\ac{rvpp} profit affected by \ac{srm} price uncertainty during period $t$ \nomunit{[€]}}
\nomenclature[V, 01]{$y_{r(d),t}$}{\ac{rvpp} profit (cost) affected by \ac{ndrs} $r$ production (demand $d$) uncertainty during period $t$ \nomunit{[€]}}
\nomenclature[V, 02]{$\eta_t^{(\prime)\rm DA}$}{Dual variable to model the negative (positive) price uncertainty of \ac{dam} during period $t$ \nomunit{[€]}}
\nomenclature[V, 02]{$\eta_t^{\rm SR} $}{Dual variable to model the price uncertainty of \ac{srm} during period $t$ \nomunit{[€]}}
\nomenclature[V, 02]{$\eta_{r(d),t}$}{Dual variable to model the \ac{ndrs} $r$ production (demand $d$) uncertainty during period  $t$ \nomunit{[€]}}
\nomenclature[V, 02]{$\nu^{\rm{DA/SR}} $}{Dual variable to model the price uncertainty of \ac{dam}/\ac{srm} \nomunit{[€]}}
\nomenclature[V, 02]{$\nu_{r(d)}$}{Dual variable to model the \ac{ndrs} $r$ production (demand $d$) uncertainty during period $t$ \nomunit{[€]}}\vspace{5pt}

\nomenclature[Z, 01]{$u_{d,p}$}{Indicator of selection of profile $p$ of demand $d$ \nomunit{[-]}}
\nomenclature[Z, 02]{$\chi_{r(d),t}$}{Binary variable that is 1 if \ac{ndrs} $r$ (demand $d$) robust constraints are active during period $t$, and 0 otherwise \nomunit{[-]}\vspace{5pt}}
\nomenclature[Z, 02]{$\chi_t^{(\prime) DA}$}{Binary variable that is 1 if \ac{dam} negative (positive) price robustness constraints are active during period $t$, and 0 otherwise \nomunit{[-]}}
\nomenclature[Z, 02]{$\chi_t^{SR}$}{Binary variable that is 1 if \ac{srm} price robustness constraints are active during period $t$, and 0 otherwise \nomunit{[-]}}

\renewcommand{\nomname}{}
\printnomenclature[1.5cm]

\vspace{-5mm}
\subsection{Price Robustness (Objective Function)}

The objective function of simultaneous \ac{rvpp} participation in the \ac{dam} and \ac{srm}, as well as the associated robust constraints, are presented in this section. 

The objective function~\eqref{obj: RTO} maximizes the benefits of \ac{rvpp} in the \ac{dam} and \ac{srm}. The first and second lines of~\eqref{obj: RTO} calculate the expected \ac{rvpp} incomes from bidding in the \ac{dam} and from up and down SR provision, respectively, considering the corresponding robustness cost of asymmetric electricity price uncertainties. 
The third line in~\eqref{obj: RTO} defines the operation costs of \acp{ndrs}, and the costs of selecting a particular load profile. 
Note that the implementation of variables $y_t^{\rm DA}$, $y_t^{\prime DA}$, $y_t^{\rm SR,\uparrow}$, and $y_t^{\rm SR,\downarrow}$ in the objective function is one of the main differences between the proposed model and the common approach to model the price robustness in the literature~\cite{rahimiyan2015strategic,Hadi}. By means of these variables, the final value of the \ac{dam}/\ac{srm} electricity price and the traded energy/reserve of \ac{rvpp} are used to calculate the worst-case scenarios.

\begin{multline} \label{obj: RTO}
\mathop {\max }\limits_{{\Xi ^{\rm DA+SR}}} \sum\limits_{t \in \mT} {\left[ {\tilde \lambda _t^{\rm DA}p_t^{\rm DA}\Delta t - y_t^{\rm{DA}} - y_t^{\prime \DA}} \right]} \\
+ \sum\limits_{t \in \mT} {\left[ {\tilde \lambda _t^{\rm{SR},\uparrow}r_t^{\SR,\uparrow} +\tilde \lambda _t^{\SR,\downarrow}r_t^{\SR,\downarrow} - y_t^{\SR,\uparrow} - y_t^{\SR,\downarrow}} \right]} \\
- \sum\limits_{t \in \mT} {\sum\limits_{r \in \mR} {C_r^Rp_{r,t}^{\DA}\Delta t} }  - \sum\limits_{d \in \mD} {\sum\limits_{p \in \mP} {{C_{d,p}}{u_{d,p}}} }  \hspace{3.7em}
\end{multline}

The set of constraints~\eqref{cons: obj} is related to the uncertainties of the \ac{dam} electricity price and are written by developing the approach in~\cite{bertsimas2004price, rahimiyan2015strategic, Hadi} and elaborating on the big M method~\cite{floudas1995nonlinear}. 
\begin{subequations}
\begin{IEEEeqnarray}{llr}
     \lambda _t^{DA} = \tilde \lambda _t^{DA} - \check{\lambda}_t^{DA}\chi _t^{DA} + \hat \lambda _t^{DA}\chi_t^{\prime {DA}}~, 
     & 
      \forall t \qquad  \label{cons: obj1} \\     
    {\nu ^{DA}} + \eta _t^{DA} \ge  \check{\lambda}_t^{DA}p_t^{DA}\Delta t\,~, 
    &   
    \forall t \label{cons: obj2} \\ 
    {\nu ^{DA}} + \eta_t^{\prime {DA}} \ge  - \hat \lambda _t^{DA}p_t^{DA}\Delta t\,~, 
    &   
    \forall t  \label{cons: obj3} \\ 
    y_t^{DA} \ge {\nu^{DA}} + \eta_t^{DA} - M(1 - \chi_t^{DA})\,~, 
    &  
    \forall t \label{cons: obj4}  \\ 
    y_t^{\prime {DA}} \ge {\nu ^{DA}} + \eta_t^{\prime {DA}} - M(1 - \chi_t^{\prime {DA}})~, 
    &    
    \forall t  \label{cons: obj5} \\
    \varepsilon (\chi _t^{DA}) \le \eta _t^{DA} \le M(\chi _t^{DA})\,~, 
    & 
    \forall t  \label{cons: obj6} \\ 
    \varepsilon (\chi_t^{\prime {DA}}) \le \eta_t^{\prime DA} \le M(\chi_t^{\prime {DA}})\,~, 
    &  
    \forall t \label{cons: obj7} \\
    - M(1 - \chi _t^{DA})\, \le \check{\lambda}_t^{DA}p_t^{DA}\Delta t - {\nu ^{DA}} \le M(\chi _t^{DA})\,~,
     & 
    \forall t  \label{cons: obj8} \\
    -M(1-\chi_t^{\prime {DA}}) \le -\hat \lambda_t^{DA}p_t^{DA}\Delta t-{\nu^{DA}} \le M(\chi_t^{\prime {DA}})
     & 
     \forall t  \label{cons: obj9} \\
    \sum\limits_{t \in \mT} {(\chi _t^{DA} + \chi_t^{\prime {DA}})}  = {\Gamma ^{DA}}\,~,
    &  
    \phantom{\forall t}  \label{cons: obj10} \\
    \chi _t^{DA} + \chi_t^{\prime {DA}} \le 1\,~,
    &  
    \forall t  \label{cons: obj11}  \\  
    \nu^{DA},\eta_t^{DA},\eta_t^{\prime DA},y_t^{DA},y_t^{\prime DA} \ge 0\,~, 
    & 
    \forall t  \label{cons: obj12}  \\ 
    \chi _{t}^{DA},\chi _{t}^{\prime DA} \in \left\{ {0,1} \right\}\,~, 
    &  
    \forall t  \label{cons: obj13}
\end{IEEEeqnarray}
\label{cons: obj}
\end{subequations}

\vspace{-4mm}
Constraint~\eqref{cons: obj1} determines the \ac{dam} electricity price in each time period according to the condition of binary variables $\chi _t^{\DA}$ and $\chi_t^{\prime {\DA}}$, which are related to the negative and positive price volatility, respectively. Constraints~\eqref{cons: obj2} and~\eqref{cons: obj3} model the impact of the absolute value of negative and positive price volatility on profit reduction when the electricity price takes its worst condition. Constraints~\eqref{cons: obj4} and~\eqref{cons: obj5} set a lower bound for the profit reduction variables $y_t^{\DA}$ and $y_t^{\prime {\DA}}$ due to the negative and positive price uncertainty, respectively. When the binary variable $\chi_t^{\DA}$ $\left(\chi_t^{\prime {\DA}}\right)$ is 1, constraint~\eqref{cons: obj4} (constraint~\eqref{cons: obj5}) is active. Depending on whether \ac{rvpp} sells or buys electricity on the market, the worst \ac{dam} price conditions occur at the price values $\tilde \lambda _t^{\DA}-\check{\lambda}_t^{\DA}$ and $\tilde \lambda _t^{DA}+\hat{\lambda}_t^{\DA}$, respectively. The dual variables $\eta _t^{DA}$ and $\eta_t^{\prime {\DA}}$, related to the negative and positive deviations of the electricity price, are logically constrained by~\eqref{cons: obj6} and~\eqref{cons: obj7}, respectively, based on the active or non-active status of the periods to comply with the robustness budget defined in~\eqref{cons: obj10}. Constraints~\eqref{cons: obj8} and~\eqref{cons: obj9} define the lower and upper bounds for the differences between the possible profit reductions (due to the negative price deviation $\check{\lambda}_t^{DA}p_t^{DA}\Delta t$ and the positive price deviation $-\hat \lambda _t^{DA}p_t^{DA}\Delta t$) and the dual variable ${\nu^{\DA}}$. According to these constraints, the possible profit reductions must be greater than or equal to the dual variable ${\nu ^{\DA}}$ for those periods that the electricity price fluctuates to its worst case. Constraints~\eqref{cons: obj8} and~\eqref{cons: obj9} are thus essential to avoid selecting incorrect periods for the worst case of profit deviations, especially when other uncertain parameters such as ND-RESs production and demand (see Sections~\ref{sub:ndrs_profit_Robustness} and~\ref{sub:demand_Cost_Robustness}) affect the total power traded by \ac{rvpp} ($p_t^{DA}$). The robustness budget $\Gamma ^{DA}$ in~\eqref{cons: obj10} is a user-defined parameter that determines the number of periods in which the electricity price can deviate to its worst condition. $\Gamma ^{DA}$ is thus a particularly relevant parameter in this model. Constraint~\eqref{cons: obj11} prevents positive and negative electricity price deviations in the same period. Constraints~\eqref{cons: obj12} and~\eqref{cons: obj13} define the nature of positive dual variables and binary variables, respectively. The use of two binary variables to define the negative and positive price deviations and the implementation of constraints~\eqref{cons: obj8} and~\eqref{cons: obj9} to avoid illogical condition for the price deviations is another significant improvement compared to previous robust formulations in the literature~\cite{rahimiyan2015strategic,Hadi}.

The set of constraints~\eqref{cons: obj_reserve} related to the uncertainty in the up and down \ac{srm} price is defined similarly to~\eqref{cons: obj}. The only difference is that for both up and down \ac{srm} price, only the negative \ac{srm} price deviations due to uncertainty are considered in~\eqref{cons: obj_reserve1} and~\eqref{cons: obj_reserve2}, respectively. This is due to the fact that the positive \ac{srm} price deviations always result in more benefit for \ac{rvpp}. Therefore, the maximum possible profit deviations $\check{\lambda}_t^{SR,\uparrow}r_t^{SR,\uparrow}$ and $\check \lambda _t^{SR,\downarrow}r_t^{SR,\downarrow}$ are calculated based on the negative upward and downward \ac{srm} price deviations $\check{\lambda}_t^{SR,\uparrow}$ and $\check \lambda _t^{SR,\downarrow}$ in constraints~\eqref{cons: obj_reserve3} and~\eqref{cons: obj_reserve4}, respectively.
\begin{subequations}
\begin{IEEEeqnarray}{llr}
    \lambda _t^{SR,\uparrow} = \tilde \lambda _t^{SR,\uparrow} - \check{\lambda}_t^{SR,\uparrow}\chi _t^{SR,\uparrow}~, 
    &
    \forall t \qquad \label{cons: obj_reserve1} \\
    \lambda _t^{SR,\downarrow} = \tilde \lambda _t^{SR,\downarrow} - \check{\lambda}_t^{SR,\downarrow}\chi _t^{SR,\downarrow}~, 
    & 
    \forall t  \label{cons: obj_reserve2} \\
    {\nu ^{SR,\uparrow}} + \eta _t^{SR,\uparrow} \ge  \check{\lambda}_t^{SR,\uparrow}r_t^{SR,\uparrow}~, 
    &
    \forall t  \label{cons: obj_reserve3} \\
    {\nu ^{SR,\downarrow}} + \eta_t^{SR,\downarrow} \ge 
    \check \lambda _t^{SR,\downarrow}r_t^{SR,\downarrow}~, 
    &
    \forall t \label{cons: obj_reserve4} \\
    y_t^{SR,\uparrow} \ge {\nu^{SR,\uparrow}} + \eta_t^{SR,\uparrow} - M(1 - \chi_t^{SR,\uparrow})~,
    &
    \forall t \label{cons: obj_reserve5}  \\
    y_t^{{SR,\downarrow}} \ge {\nu ^{SR,\downarrow}} + \eta_t^{{SR,\downarrow}} - M(1 - \chi_t^{{SR,\downarrow}})~, 
    &
    \forall t  \label{cons: obj_reserve6} \\
    \varepsilon (\chi _t^{SR,\uparrow}) \le \eta _t^{SR,\uparrow} \le M(\chi _t^{SR,\uparrow})~, 
    &
    \forall t  \label{cons: obj_reserve7} \\
    \varepsilon (\chi_t^{SR,\downarrow}) \le \eta_t^{SR,\downarrow} \le M(\chi_t^{SR,\downarrow})~, 
    & 
    \forall t \label{cons: obj_reserve8} \\
    - M(1 - \chi _t^{SR,\uparrow})\, \le \check{\lambda}_t^{SR,\uparrow}r_t^{SR,\uparrow} - {\nu ^{SR,\uparrow}} \le M(\chi _t^{SR,\uparrow})
    &
    \forall t \label{cons: obj_reserve9} \\
    - M(1 - \chi_t^{SR,\downarrow})\, \le  \check \lambda _t^{SR,\downarrow}r_t^{SR,\downarrow} - {\nu ^{SR,\downarrow}} \le M(\chi_t^{SR,\downarrow})
    &
    \forall t \label{cons: obj_reserve10} \\
    \sum\limits_{t \in \mT} {\chi _t^{SR,\uparrow}}  = {\Gamma ^{SR,\uparrow}}~, 
    &
    \phantom{\forall t} \label{cons: obj_reserve11} \\
    \sum\limits_{t \in \mT} {\chi_t^{SR,\downarrow}}  = {\Gamma ^{SR,\downarrow}}~, 
    &
    \phantom{\forall t} \label{cons: obj_reserve12} \\
    \nu^{SR,\uparrow}, \nu^{SR,\downarrow}, \eta_t^{SR,\uparrow}, \eta_t^{SR,\downarrow}, y_t^{SR,\uparrow}, y_t^{SR,\downarrow} \ge 0~, 
    &
    \forall t  \label{cons: obj_reserve13}  \\
    \chi _{t}^{SR,\uparrow},\chi _{t}^{SR,\downarrow} \in \left\{ {0,1} \right\}~, 
    &
    \forall t \label{cons: obj_reserve14}
\end{IEEEeqnarray}
\label{cons: obj_reserve}
\end{subequations}
\vspace{-4mm}

\subsection{Supply-demand and Traded Constraints}

The supply-demand balancing constraint of \ac{rvpp} units is defined in~\eqref{cons: Supply-Demand}. All \ac{rvpp} units are assumed to be connected to a single node. The variable $r_{t}^{SR}$ related to the total traded reserve of \ac{rvpp} and the variables $r_{r,t}^{SR}$ and $r_{d,t}^{SR}$ related to the reserve of \ac{rvpp} units are defined according to different reserve activation scenarios similar to~\cite{Hadi}. Constraints~\eqref{cons: Power-Traded1} and~\eqref{cons: Power-Traded2} assign the upper and lower bounds of traded energy and reserve by \ac{rvpp}, respectively. Constraint~\eqref{cons: Power-Traded3} defines the proportion of down and up reserve requested by \ac{tso}. The up reserve provided is limited by~\eqref{cons: Power-Traded4} to a fraction of the total capacity of the generating units of \ac{rvpp}.
\vspace{2mm}
\begin{subequations}
\begin{IEEEeqnarray}{llr}
    \sum\limits_{r \in \mR} ({p_{r,t}^{DA}} + r_{r,t}^{SR}) - \sum\limits_{d \in \mD} ({p_{d,t}^{DA}} - r_{d,t}^{SR}) = p_{t}^{DA} + r_{t}^{SR}~,
    &
    \forall t \qquad \label{cons: Supply-Demand} \\
    p_{t}^{DA} + r_{t}^{SR,\uparrow} \le \sum\limits_{r \in \mR} {\bar P_r}~, 
    &
    \forall t \label{cons: Power-Traded1} \\
    -\sum\limits_{d \in \mD} {\bar P_d} \le p_{t}^{DA} - r_{t}^{SR,\downarrow}~, 
    &
    \forall t \label{cons: Power-Traded2} \\
    r_{t}^{SR,\uparrow} = \varrho_t r_{t}^{SR,\downarrow}~, 
    &
    \forall t \label{cons: Power-Traded3} \\
    r_{t}^{SR,\uparrow} \le \kappa \sum\limits_{r \in \mR} {\bar P_r}~, 
    &
    \forall t \label{cons: Power-Traded4} 
\end{IEEEeqnarray}
\label{cons: Power-Traded}
\end{subequations}

\vspace{-7mm}

\subsection{\ac{ndrs} Profit Robustness}
\label{sub:ndrs_profit_Robustness}
The profit robustness formulation of \acp{ndrs} is given in~\eqref{cons: NDRES}. Constraint~\eqref{cons: NDRES1} is the lower bound on the \acp{ndrs} output power. Constraint~\eqref{cons: NDRES2} sets the \acp{ndrs} output through the median forecast generation of \acp{ndrs} $\tilde P_{r,t}$ and the possible negative power deviation $\chi _{r,t}\check{P}_{r,t}$ (active when $\chi _{r,t}=1$). 
The binary variable $\chi _{r,t}$ is determined according to the profit robustness constraints~\eqref{cons: NDRES3}-\eqref{cons: NDRES10} proposed in this work. 
Constraint~\eqref{cons: NDRES3} limits the profit of \acp{ndrs} for each time period by considering the robustness of the problem against uncertain parameters of \acp{ndrs} production. The upper bound of this constraint is computed as the median profit minus the profit reduction due to the negative deviation of power forecast of \acp{ndrs}, $y_{r,t}$. The median profit is calculated by multiplying the electricity price $\lambda _t^{DA}$ and the median production of \acp{ndrs} minus the provided up reserve, both multiplied by the time period duration $(\tilde P_{r,t}-r_{r,t}^{SR,\uparrow})\Delta t$. Constraint~\eqref{cons: NDRES3} is a non-linear expression that is linearized in Section~\ref{sub:non-linear_constraints}. Constraint~\eqref{cons: NDRES4} assigns the upper bound of the dual variable $y_{r,t}$ to the negative profit deviation $\lambda _t^{DA}\check{P}_{r,t}\Delta t$ of each \acp{ndrs} due to uncertainty. To model the worst-case scenarios of profit reduction for each \ac{ndrs}, only negative power deviations are considered in this constraint, since positive deviations will usually benefit the \ac{rvpp}. Constraint~\eqref{cons: NDRES5} determines the lower bound of the dual variable $y_{r,t}$ according to the dual variables $\nu _r$ and $\eta _{r,t}$, and the condition of the binary variable $\chi _{r,t}^{DA}$. Constraint~\eqref{cons: NDRES6} assigns the lower bound of the sum of the dual variables $\nu _r$ and $\eta _{r,t}$ to the maximum profit reduction for each \ac{ndrs} in each time period. According to constraint~\eqref{cons: NDRES7}, the dual variable $\eta _{r,t}$ is defined based on the active or non-active status of the profit reduction due to the robustness of the production of \acp{ndrs}. Constraint~\eqref{cons: NDRES8} defines the profit robustness budget for each \ac{ndrs}.
%
\begin{subequations}
{\allowdisplaybreaks\begin{IEEEeqnarray}{llr}
    \ubar P_{r} \le p_{r,t}^{DA}-r_{r,t}^{SR,\downarrow}~, 
    & & \quad
    \forall r,t \qquad \label{cons: NDRES1} \\ 
    p_{r,t}^{DA}+r_{r,t}^{SR,\uparrow} = \tilde P_{r,t} - \chi _{r,t}\check{P}_{r,t}~, 
    & &  \quad
    \forall r,t \qquad \label{cons: NDRES2} \\
    \lambda _t^{DA}p_{r,t}^{DA}\Delta t \le \lambda _t^{DA} (\tilde P_{r,t} - r_{r,t}^{SR,\uparrow})\Delta t - y_{r,t}~,
    & &  \quad
    \forall r,t \qquad \label{cons: NDRES3} \\
    y_{r,t} \le \lambda _t^{DA}\check{P}_{r,t}\Delta t~, 
    & &  \quad
    \forall r,t \qquad \label{cons: NDRES4} \\
    y_{r,t} \ge \nu _r + \eta _{r,t} - M(1 - \chi _{r,t})~,
    & &  \quad
    \forall r,t \qquad \label{cons: NDRES5} \\
    \nu _r + \eta _{r,t} \ge \lambda _t^{DA}\check{P}_{r,t}\Delta t~, 
    & &  \quad
    \forall r,t \qquad \label{cons: NDRES6} \\ 
    \varepsilon \chi _{r,t} \le \eta _{r,t} \le M\chi _{r,t}~, 
    & &  \quad 
    \forall r,t \qquad \label{cons: NDRES7} \\ 
    \sum\limits_{t \in \mT} {\chi _{r,t}}  = \Gamma _r~, 
    & &  \quad
    \forall r\phantom{,t}  \qquad \label{cons: NDRES8} \\ 
    \nu _r,\eta _{r,t},y_{r,t} \ge 0~, 
    & &  \quad
    \forall r,t \qquad \label{cons: NDRES9} \\ 
    \chi _{r,t} \in \left\{ {0,1} \right\}~, 
    & &  \quad
    \forall r,t \qquad \label{cons: NDRES10}
\end{IEEEeqnarray}}
\label{cons: NDRES}
\end{subequations}
\vspace{-3em}
\subsection{Demand Cost Robustness}
\label{sub:demand_Cost_Robustness}
The demand cost robust formulation is illustrated in~\eqref{cons: Demand}, which is based on the deterministic model presented in~\cite{oladimeji2022modeling}. Constraint~\eqref{cons: Demand1} assigns the demand for each period to predefined demand profiles, taking into account the median and positive demand forecasts. Only positive demand deviations are considered for the worst-case cost robustness scenarios, since the negative demand deviations (i.e., lower consumption) usually result in lower costs for \ac{rvpp}. Constraint~\eqref{cons: Demand2} ensures that the algorithm selects only one demand profile among several profiles. When the binary variable $\chi _{d,t}$ in~\eqref{cons: Demand1} for a certain period is 1, the possible positive deviation of the demand becomes active. The binary variable $\chi _{d,t}$ is determined according to the cost robustness constraints~\eqref{cons: Demand3}-\eqref{cons: Demand8} proposed in this work. Constraint~\eqref{cons: Demand3} sets the lower bound on the cost of buying electricity from \ac{dam} to supply demand, which equals the cost of buying electricity for the median demand forecast $\lambda _t^{DA}\sum\limits_{p \in \mP} {(\tilde P_{d,p,t} u_{d,p})}\Delta t$ plus the additional cost of positive demand fluctuation due to uncertainty represented by the dual variable $y_{d,t}$. The additional cost of buying electricity for positive demand fluctuation due to uncertainty $\lambda _t^{DA}\sum\limits_{p \in \mP} {(\hat P_{d,p,t} u_{d,p})}\Delta t$ is assigned as the upper bound of the dual variable $y_{d,t}$ by constraint~\eqref{cons: Demand4}. On the other hand, the lower bound of the dual variable $y_{d,t}$ is given by constraint~\eqref{cons: Demand5} to find the worst cases of demand cost robustness. The dual variables $\nu _d$ and $\eta _{d,t}$ are logically constrained in~\eqref{cons: Demand6} and~\eqref{cons: Demand7} to determine those periods that positive demand deviations lead to the worst cost robustness scenarios. Constraint~\eqref{cons: Demand8} assigns the user-defined parameter of the robustness budget $\Gamma _d$ to set the number of periods allowed for positive deviations in demand due to cost robustness. Constraints~\eqref{cons: Demand9} and~\eqref{cons: Demand10} confine the demand up reserve according to the percentage of downward demand flexibility and the minimum possible demand, respectively. Constraints~\eqref{cons: Demand11} and~\eqref{cons: Demand12} are similarly defined to limit the down reserve considering the opposite direction of demand flexibility and the maximum possible demand. The worst conditions of ramp-up and ramp-down in two consecutive periods considering the reserve activation are defined in constraints~\eqref{cons: Demand13} and~\eqref{cons: Demand14}, respectively. The capability of demand to provide up and down reserve is defined by constraints~\eqref{cons: Demand15} and~\eqref{cons: Demand16}, respectively. Constraint~\eqref{cons: Demand17} limits the minimum energy that each demand should use for the entire period. Constraints~\eqref{cons: Demand18} and~\eqref{cons: Demand19} describe the nature of positive dual variables and binary variables, respectively.
\vspace{-.25em}
\begin{subequations}
{\allowdisplaybreaks\begin{IEEEeqnarray}{llr}
    p_{d,t}^{DA} = \sum\limits_{p \in \mP} {(\tilde P_{d,p,t} u_{d,p}
    + \chi_{d,t}\hat P_{d,p,t}u_{d,p})}~,
    & & \quad
    \forall d,t  \qquad \label{cons: Demand1}\\
    \sum\limits_{P \in \mP} {u_{d,p}} = 1~, 
    & & \quad
    \forall d\phantom{,t} \qquad \label{cons: Demand2}\\
    \lambda _t^{DA}p_{d,t}^{DA}\Delta t \ge \lambda _t^{DA}\sum\limits_{p \in \mP} {(\tilde P_{d,p,t} u_{d,p})}\Delta t + y_{d,t}~,
    & & \quad
    \forall d,t \qquad \label{cons: Demand3}\\
    y_{d,t} \le \lambda_{t}^{DA} \sum\limits_{p \in \mP} {(\hat P_{d,p,t} u_{d,p})}\Delta t~,
    & & \quad
    \forall d,t \qquad  \label{cons: Demand4} \\
    y_{d,t} \ge \nu _d + \eta _{d,t} - M(1 - \chi _{d,t})~, 
    & & \quad
    \forall d,t \qquad  \label{cons: Demand5} \\ 
    \nu _d + \eta _{d,t} \ge \lambda _t^{DA}\sum\limits_{p \in P} {(\hat P_{d,p,t} u_{d,p})}\Delta t~, 
    & & \quad
    \forall d,t \qquad  \label{cons: Demand6} \\ 
    \varepsilon \chi _{d,t} \le \eta _{d,t} \le M\chi _{d,t}~, 
    & & \quad
    \forall d,t \qquad \label{cons: Demand7} \\ 
    \sum\limits_{t \in \mT} {\chi _{d,t}}  = \Gamma _d~, 
    & & \quad
    \forall d\phantom{,t} \qquad \label{cons: Demand8} \\ 
    r_{d,t}^{SR,\uparrow} \le \ubar\beta_{d,t} \sum\limits_{p \in \mP} {(\tilde P_{d,p,t} u_{d,p})}~,
    & & \quad
    \forall d,t \qquad \label{cons: Demand9} \\
    r_{d,t}^{SR,\uparrow} \le p_{d,t}^{DA} - \ubar P_{d}~,
    & & \quad
    \forall d,t \qquad \label{cons: Demand10} \\
    r_{d,t}^{SR,\downarrow} \le \bar\beta_{d,t} \sum\limits_{p \in \mP} {(\tilde P_{d,p,t} u_{d,p})}~,
    & & \quad
    \forall d,t \qquad \label{cons: Demand11} \\
    r_{d,t}^{SR,\downarrow} \le \bar P_{d} - p_{d,t}^{DA}~,
    & & \quad
    \forall d,t \qquad \label{cons: Demand12}\\
    (p_{d,t}^{DA} + r_{d,t}^{SR,\downarrow})- (p_{d,(t-1)}^{DA} - r_{d,(t-1)}^{SR,\uparrow})
    \le \bar R_{d} \Delta t~,
    & & \quad
    \forall d,t \qquad \label{cons: Demand13}\\
    (p_{d,(t-1)}^{DA} + r_{d,(t-1)}^{SR,\downarrow}) - (p_{d,t}^{DA} - r_{d,t}^{SR,\uparrow}) \le \ubar R_{d} \Delta t~,
    & & \quad
    \forall d,t \qquad \label{cons: Demand14}\\
    r_{d,t}^{SR,\uparrow} \le T^{SR} \ubar R_{d}^{SR}~,
    & & \quad
    \forall d,t \qquad \label{cons: Demand15} \\
    r_{d,t}^{SR, \downarrow} \le T^{SR} \bar R_{d}^{SR}~,
    & & \quad
    \forall d,t \qquad \label{cons: Demand16} \\
    \ubar E_{d} \le \sum\limits_{t \in \mT} ({p_{d,t}^{DA}\Delta t} - r_{d,t}^{SR,\uparrow})~,
    & & \quad
    \forall d\phantom{,t} \qquad \label{cons: Demand17}\\
    \nu _d,\eta _{d,t},y_{d,t} \ge 0~, 
    & & \quad
    \forall d,t \qquad \label{cons: Demand18} \\ 
    \chi _{d,t} \in \left\{ {0,1} \right\}~, 
    & & \quad
    \forall d,t \qquad \label{cons: Demand19} 
\end{IEEEeqnarray}}
\label{cons: Demand}
\end{subequations}
\vspace{-3em}
\subsection{Coping with Non-linear Constraints}
\label{sub:non-linear_constraints}
This section discusses the derivation from the non-linear terms in sets of equations~\eqref{cons: NDRES} and~\eqref{cons: Demand} to obtain a single-level \ac{milp} problem with an exact solution. The set of equations~\eqref{cons: NDRES} contains two non-linear terms on the left and right hand sides of~\eqref{cons: NDRES3} due to the multiplication of the electricity price variable $\lambda _t^{DA}$ and the continuous variables $p_{r,t}^{DA}$ and $r_{r,t}^{SR,\uparrow}$. By substituting the electricity price from constraint~\eqref{cons: obj1} into the non-linear terms, the profit robustness constraint~\eqref{cons: NDRES3} can be rewritten as~\eqref{cons: NDRES_Linear1}. In equation~\eqref{cons: NDRES_Linear1}, the non-linear terms $\check{\lambda}_t^{DA}\chi_t^{DA} (p_{r,t}^{DA}+r_{r,t}^{SR,\uparrow})\Delta t$ and $\hat \lambda _t^{DA}\chi_t^{\prime {DA}} (p_{r,t}^{DA}+r_{r,t}^{SR,\uparrow})\Delta t$ are the multiplications of binary and continuous variables. Note that the consideration of discrete rather than continuous values for the electricity price in~\eqref{cons: obj1} is relevant to the robustness concept, since the worst-case scenarios occur in the boundary values of electricity price. Finally, the non-linear equation~\eqref{cons: NDRES_Linear1} can be replaced by the set of linear constraints~\eqref{cons: NDRES_Linear2}-\eqref{cons: NDRES_Linear8} using the method in~\cite{floudas1995nonlinear}.
\vspace{-.5em}
\begin{subequations}
\begin{multline}
     \hspace{1mm} \tilde \lambda _t^{DA} (p_{r,t}^{DA}+r_{r,t}^{SR,\uparrow})\Delta t \\
    - \check{\lambda}_t^{DA}\chi _t^{DA} (p_{r,t}^{DA}+r_{r,t}^{SR,\uparrow})\Delta t \\
    + \hat \lambda _t^{DA}\chi_t^{\prime {DA}} (p_{r,t}^{DA}+r_{r,t}^{SR,\uparrow})\Delta t \\
    \le \lambda _t^{DA}\tilde P_{r,t}\Delta t - y_{r,t}~, 
    \quad \forall r,t  \label{cons: NDRES_Linear1} 
\end{multline}
\vspace{-2em}
\begin{multline}
     \hspace{1mm} \tilde \lambda _t^{DA} (p_{r,t}^{DA}+r_{r,t}^{SR,\uparrow})\Delta t - \check{\lambda}_t^{DA}p_{r,t}^{DA,Q}\Delta t \\ + \hat \lambda _t^{DA}p_{r,t}^{\prime DA,Q}\Delta t \le
    \lambda _t^{DA}\tilde P_{r,t}\Delta t - y_{r,t}~,
     \quad \forall r,t  \label{cons: NDRES_Linear2}
\end{multline}
\vspace{-2em}
\begin{IEEEeqnarray}{llr}
    p_{r,t}^{DA,Q} = p_{r,t}^{DA}+r_{r,t}^{SR,\uparrow} - p_{r,t}^{DA,A}~,
    & & \quad
    \forall r,t \quad \label{cons: NDRES_Linear3}\\
    \ubar P_{r}\chi _t^{DA} \le p_{r,t}^{DA,Q} \le \tilde P_{r,t}\chi _t^{DA}~,
    & & \quad 
    \forall r,t \quad \label{cons: NDRES_Linear4} \\
    \ubar P_{r}(1 - \chi _t^{DA}) \le p_{r,t}^{DA,A} \le \tilde P_{r,t}(1 - \chi _t^{DA})~, 
    & & \quad
    \forall r,t \quad \label{cons: NDRES_Linear5} \\
    p_{r,t}^{\prime DA,Q} = p_{r,t}^{DA}+r_{r,t}^{SR,\uparrow} - p_{r,t}^{\prime DA,A}~,
    & & \quad    
    \forall r,t \quad \label{cons: NDRES_Linear6} \\
    \ubar P_{r}\chi_t^{\prime {DA}} \le p_{r,t}^{\prime DA,Q} \le \tilde P_{r,t}\chi_t^{\prime {DA}}~,
     & & \quad
    \forall r,t \quad \label{cons: NDRES_Linear7} \\
    \ubar P_{r}(1 - \chi_t^{\prime {DA}}) \le p_{r,t}^{\prime DA,A} \le \tilde P_{r,t}(1 - \chi_t^{\prime {DA}})~,
    & & \quad
    \forall r,t \quad \label{cons: NDRES_Linear8} 
\end{IEEEeqnarray}
\label{cons: NDRES_Linear}
\end{subequations}
The auxiliary variables $p_{r,t}^{DA,Q}$ and $p_{r,t}^{DA,A}$ with the same possible lower and upper bounds as the term $p_{r,t}^{DA}+r_{r,t}^{SR,\uparrow}$ are defined to determine the final result of the non-linear term $\check{\lambda}_t^{DA}\chi_t^{DA} (p_{r,t}^{DA}+r_{r,t}^{SR,\uparrow}) \Delta t$. When the binary variable $\chi _t^{DA}$ related to the negative electricity price deviation is 1, equations~\eqref{cons: NDRES_Linear3}-\eqref{cons: NDRES_Linear5} set $p_{r,t}^{DA,Q}$ = $p_{r,t}^{DA}+r_{r,t}^{SR,\uparrow}$ and $p_{r,t}^{DA,A}$ = 0. On the other hand, for $\chi _t^{DA}$ = 0, equations~\eqref{cons: NDRES_Linear3}-\eqref{cons: NDRES_Linear5} lead to $p_{r,t}^{DA,Q}$ = 0 and $p_{r,t}^{DA,A}$ = $p_{r,t}^{DA}+r_{r,t}^{SR,\uparrow}$.
Similarly, the auxiliary variables $p_{r,t}^{\prime DA,Q}$ and $p_{r,t}^{\prime DA,A}$ in equations~\eqref{cons: NDRES_Linear6}-\eqref{cons: NDRES_Linear8} can define the final result of the non-linear term $\hat \lambda _t^{DA}\chi_t^{\prime {DA}} (p_{r,t}^{DA}+r_{r,t}^{SR,\uparrow})\Delta t$ in~\eqref{cons: NDRES_Linear1}. Therefore, the linear equations~\eqref{cons: NDRES_Linear2}-\eqref{cons: NDRES_Linear8} can replace the non-linear constraint~\eqref{cons: NDRES_Linear1}.

The demand robust cost formulation proposed in~\eqref{cons: Demand} includes non-linear terms in~\eqref{cons: Demand1},~\eqref{cons: Demand3},~\eqref{cons: Demand4}, and~\eqref{cons: Demand6}. The non-linear term $\lambda _t^{DA}p_{d,t}^{DA}\Delta t$ in~\eqref{cons: Demand3} can be linearized in the same way as in~\eqref{cons: NDRES_Linear} by introducing new auxiliary variables. In addition, each of the non-linear terms $\sum\limits_{p \in \mP} {(\chi_{d,t}\hat P_{d,p,t}u_{d,p})}$ in~\eqref{cons: Demand1}, and, by including the expanded term of the electricity price $\lambda _t^{DA}$ from constraint~\eqref{cons: obj1}, the non-linear terms $\lambda _t^{DA}\sum\limits_{p \in \mP} {(\tilde P_{d,p,t} u_{d,p})}\Delta t$ in~\eqref{cons: Demand3}, and $\lambda _t^{DA}\sum\limits_{p \in \mP} {(\hat P_{d,p,t} u_{d,p})}\Delta t$ in~\eqref{cons: Demand4} and~\eqref{cons: Demand6} includes only the multiplication of two binary variables. To linearize these binary multiplication terms, three new binary variables $z_{d,p,t}$, $w_{d,p,t}$, $w_{d,p,t}^{\prime}$ are introduced as the final result of binary multiplications of $\chi_{d,t}u_{d,p}$, $\chi_{t}^{DA}u_{d,p}$, and $\chi_{t}^{\prime DA}u_{d,p}$, respectively. Furthermore, the set of linear constraints~\eqref{cons: Demand_Linear} is added to~\eqref{cons: Demand}, which simulate the possible results of multiplying two binary variables by the newly defined binary variables $z_{d,p,t}$, $w_{d,p,t}$, $w_{d,p,t}^{\prime}$.
\vspace{-.5em}
\begin{subequations}
\begin{IEEEeqnarray}{llr}
z_{d,p,t} \le \chi_{d,t}~, \hspace{35mm}
& & \qquad \quad
\forall d,p,t \qquad \, \label{cons: Demand_Linear1} \\
z_{d,p,t} \le u_{d,p}~, 
& & \quad
\forall d,p,t \qquad \, \label{cons: Demand_Linear2} \\
z_{d,p,t} + 1 \ge \chi_{d,t} + u_{d,p}~, 
& & \quad
\forall d,p,t \qquad \, \label{cons: Demand_Linear3} \\
w_{d,p,t} \le \chi_{t}^{DA}~, 
& & \quad
\forall d,p,t \qquad \, \label{cons: Demand_Linear4} \\
w_{d,p,t} \le u_{d,p}~, 
& & \quad
\forall d,p,t \qquad \, \label{cons: Demand_Linear5} \\
w_{d,p,t} + 1 \ge \chi_{t}^{DA} + u_{d,p}~, 
& & \quad
\forall d,p,t \qquad \, \label{cons: Demand_Linear6} \\
w_{d,p,t}^{\prime} \le \chi_{t}^{\prime DA}~, 
& & \quad
\forall d,p,t \qquad \, \label{cons: Demand_Linear7} \\
w_{d,p,t}^{\prime} \le u_{d,p}~, 
& & \quad
\forall d,p,t \qquad \, \label{cons: Demand_Linear8} \\
w_{d,p,t}^{\prime} + 1 \ge \chi_{t}^{\prime DA} + u_{d,p}~, 
& & \quad
\forall d,p,t \qquad \, \label{cons: Demand_Linear9}
\end{IEEEeqnarray}
\label{cons: Demand_Linear}
\end{subequations}
\vspace{-1.5em}

Finally, by substituting the linear equivalent of constraints~\eqref{cons: NDRES} and~\eqref{cons: Demand} with~\eqref{cons: NDRES_Linear} and~\eqref{cons: Demand_Linear}, problem~\eqref{obj: RTO}-\eqref{cons: Demand} can be written as an \ac{milp} problem solvable with available \ac{milp} solvers such as CPLEX.

\vspace{-2mm}

\section{Profit Robustness Example}
\label{sec: Example}
\begin{figure}
    \centering
    
    \subfigure{\includegraphics[width=0.9\columnwidth]{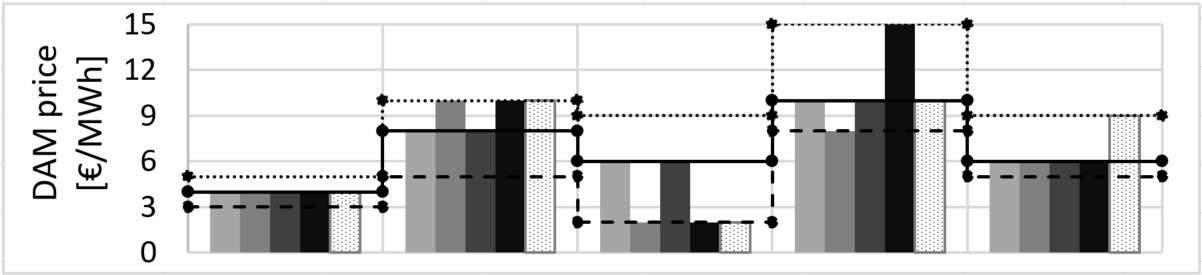}}
    \label{fig:DAM_price}

    \vspace{-3mm}
    
    \subfigure{\includegraphics[width=0.9\columnwidth]{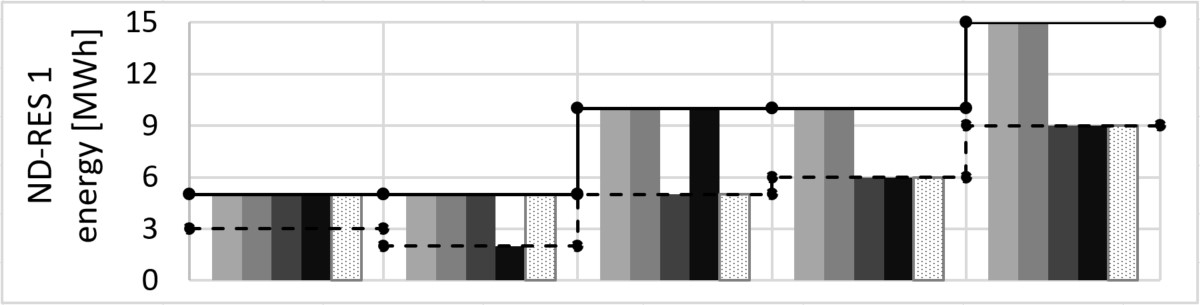}}
    \label{fig:ND1_energy}

    \vspace{-3mm}

    \subfigure{\includegraphics[width=0.9\columnwidth]{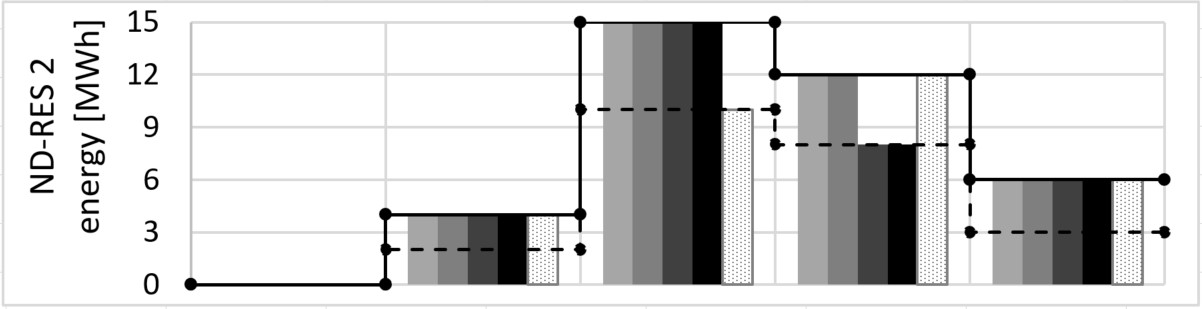}}
    \label{fig:ND2_energy}

    \vspace{-3mm}

    \subfigure{\includegraphics[width=0.9\columnwidth]{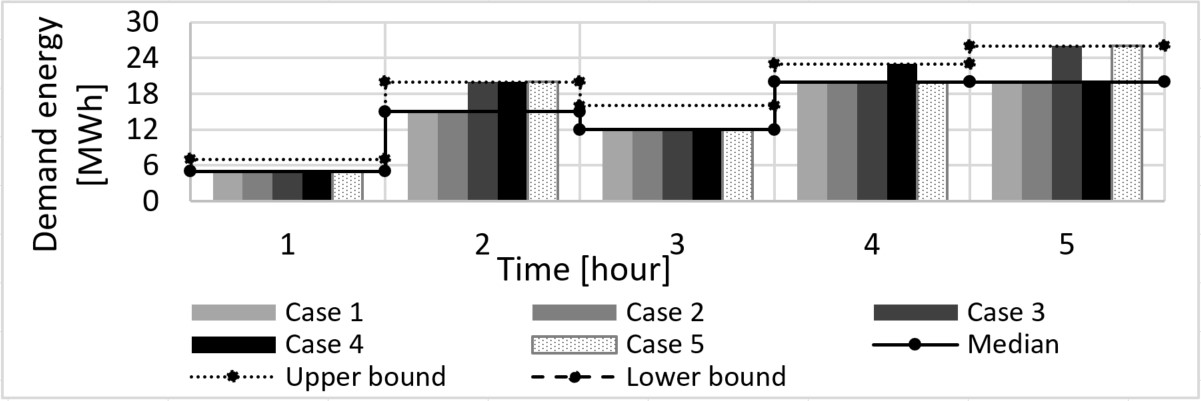}}
    \label{fig:Demand_energy}
    \vspace{-.5em}
    \caption{Final values of \ac{dam} electricity price and \ac{rvpp} units output energy in different case studies.}
    \vspace{-0.2em}
    \label{fig:Example_DAMprice_energy}
\end{figure}

This section presents a simple illustrative example to show the performance of the proposed robust formulation in finding the worst-case profit robustness scenarios by considering the asymmetry of the \ac{dam} electricity price. The example provides a detailed description of how the worst cases of the electricity price deviations affect the worst cases of energy deviations. In this context, an \ac{rvpp} with two \acp{ndrs} and one demand in a sample period of 5 hours is considered. The forecast bounds of production and demand of the \ac{rvpp} units and the \ac{dam} electricity price are shown by the dashed/solid lines in Figure~\ref{fig:Example_DAMprice_energy}. Five cases are defined below to compare different conditions for the values of energy and price uncertainty budgets and to compare the results of the proposed model by the model in~\cite{Hadi}:

\begin{itemize}
    \item Case 1: Deterministic case (i.e., $\Gamma ^{DA}=\Gamma_{r}=\Gamma_{d}=0$);
    
    \item Case 2: Only the \ac{dam} electricity price uncertainty is considered. It is assumed that the values of the \ac{dam} electricity price can deviate from the median to the worst case values in three periods (i.e., $\Gamma ^{DA}=3$ and $\Gamma_{r}=\Gamma_{d}=0$);
    
    \item Case 3: Only the uncertainty of \ac{ndrs} units energy and demand is considered. It is assumed that the production values of \ac{ndrs} 1 and \ac{ndrs} 2 and the demand can deviate from the median to the worst case values in three, one, and two periods, respectively (i.e., $\Gamma ^{DA}=0$ and $\Gamma_{r1}=3$, $\Gamma_{r2}=1$, and $\Gamma_{d}=2$);

    \item Case 4: Both price and energy uncertainties are considered. The electricity price, the production of \ac{ndrs} 1 and \ac{ndrs} 2, and the demand values can deviate from the median to the worst case values ($\Gamma ^{DA}=3$ and $\Gamma_{r1}=3$, $\Gamma_{r2}=1$, and $\Gamma_{d}=2$).

    \item Case 5: The energy robustness problem presented in~\cite{Hadi} is solved for the same uncertainty budgets as in Case 4.
\end{itemize}

\begin{figure}
    \centering
    
    \subfigure{\includegraphics[width=0.9\columnwidth]{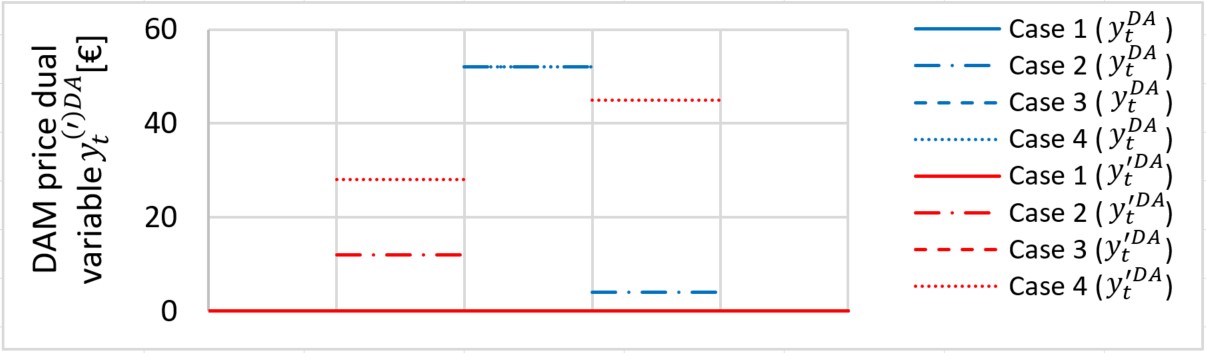}}
    \label{fig:DAM_price_Dual}

    \vspace{-3mm}
    
    \subfigure{\includegraphics[width=0.9\columnwidth]{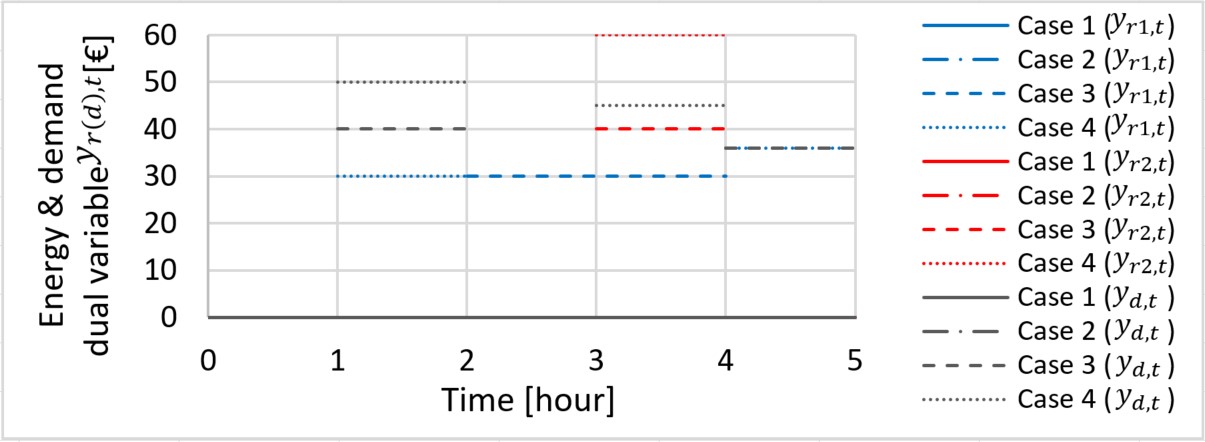}}
    \label{fig:ND1_energy_Dual}
    \vspace{-.5em}
    \caption{The profit/cost dual variables affected by different uncertainties in different case studies.}
    \vspace{-1em}
    \label{fig:Example_Dual_variables}
\end{figure}

Figure~\ref{fig:Example_DAMprice_energy} shows the final results of \ac{dam} electricity price, \acp{ndrs} energy, and demand for different cases proposed in this example. The final values of the above variables, corresponding to the whole period in each hour, are shown by different bars in this figure. If the value of a variable is equal to the median of the forecast (solid black line in the figure), it means that the corresponding period is not selected as the worst case. Figure~\ref{fig:Example_Dual_variables} shows the values of the dual variables $y_t^{(\prime) DA}$ and $y_{r(d),t}$ related to the profit/cost affected by different uncertainties for all defined cases except for Case 5. Note that, in the model in~\cite{Hadi}, these variables are either not defined or defined for energy robustness; therefore, the comparison is only provided for the first four cases.

\textit{A. Case 1:} The \ac{rvpp} obtains a profit of €56 by bidding its median values of \ac{ndrs} production according to Figure~\ref{fig:Example_DAMprice_energy}. The final values for the \ac{dam} electricity price are also obtained as the median values as the length of all bars is equal to the median. As shown in Figure~\ref{fig:Example_Dual_variables}, due to not considering the robustness, all dual variables $y_t^{(\prime) DA}$ and $y_{r(d),t}$ are equal to zero, since the problem is a deterministic optimization one.

\textit{B. Case 2:} In Case 2, the \ac{rvpp} profit in the \ac{dam} is \mbox{-€12}, where the negative value means that the cost of buying electricity to supply demand is higher than the profit obtained by selling electricity on the market. The algorithm chooses periods 3 and 4 for the negative price fluctuation and period 2 for the positive price fluctuation. Therefore, the final electricity prices in periods 3 and 4 (2) are decreased (increased) to their minimum (maximum) values compared to Case 1. 
Note that the maximum possible profit reduction in each period can be calculated by finding the maximum value of $\check{\lambda}_t^{DA}p_t^{DA}\Delta t$ for the negative price deviation and $- \hat \lambda _t^{DA}p_t^{DA}\Delta t$ for the positive price deviation. Therefore, the algorithm correctly identifies the periods that lead to the worst cases of profit reduction due to price uncertainty.

\textit{C. Case 3:} In Case 3, the \ac{rvpp} profit in the \ac{dam} is -€166. 
The maximum possible profit reduction for each period can be calculated by $\lambda _t^{DA}\check{P}_{r,t}\Delta t$ for \ac{ndrs} and the maximum possible cost increase for demand can be calculated by $\lambda _t^{DA}\hat{P}_{d,t}\Delta t$. 
The worst cases of profit reductions for \ac{ndrs} 1 occur in periods 3, 4, and 5, whereas for \ac{ndrs} 2 this occurs in period 4. The worst cases of demand cost occur in periods 2 and 5, resulting in maximum demand in these periods. It can be easily verified that the algorithm correctly selects the worst periods in terms of profit reduction for \ac{ndrs} or cost increase for demand.  

\textit{D. Case 4:} The \ac{rvpp} profit in the \ac{dam} is -€279. 
This case shows one of the significant differences between the proposed model and the models in the literature~\cite{Hadi} (by comparing the black (Case 4) and white (Case 5) bars), where instead of selecting the periods with higher energy reductions, the proposed algorithm selects the periods that result in higher profit reductions. For instance, the worst case of \ac{ndrs} 2 production occurs in period 4 with profit reduction of €60 and energy reduction of 4 MW. However, the period 3 with the highest amount of energy deviation (5 MW) in Case 5 is selected as the worst case.

In Case 5, those periods that result in more deviations of \ac{ndrs} production and demand are selected as the worst cases. Moreover, the worst cases of electricity price deviations are determined according to the final values of \ac{ndrs} production and demand. Considering the different selection of worst-case periods for Cases 4 and 5, the \ac{rvpp} obtains a profit of -€279 in the former, which is lower than the profit of -€223 obtained in Case 5. Note that the profit obtained is for bidding in the market and is different from the profit from clearing the market. Suppose the \ac{rvpp} uses the bidding strategy proposed in this paper, even though its profit is lower. In this condition, it reduces the risk of significant losses and penalties (e.g., due to buying energy in real time or penalties for the energy it promised to provide but cannot) for not considering the actual worst cases. Moreover, the results indicate that the energy robustness approach cannot fully cope with the actual worst cases for both energy (\acp{ndrs} production and demand) and price uncertainty. On the contrary, the profit robustness approach proposed in this paper considers the worst cases of profit/cost deviations for \acp{ndrs}/demand instead of the maximum energy deviation.
As a final remark, illustrative results indicate that the proposed algorithm accurately selects the worst-case profit for different uncertainty budgets, and shows better performance in finding the worst-case scenarios compared to the model in~\cite{Hadi}. These finding will be thoroughly analyzed in successive sections.

\vspace{-3mm}
\section{Simulation Results}
\label{sec:Simulation}
\begin{figure}
    \centering  \includegraphics[width=\columnwidth]{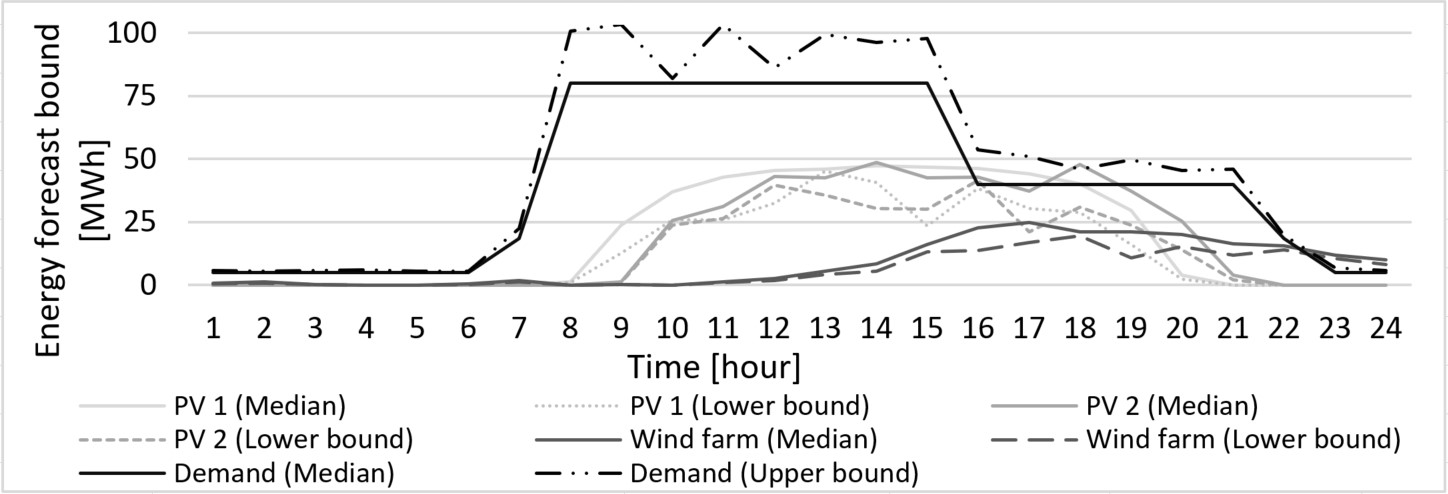}
    \vspace{-2em}
    \caption{The energy forecast data.}
    \vspace{-1.5em}
    \label{fig:Energy_forecast}
\end{figure}

This section presents the simulation results of the proposed single-level robust bidding model for different case studies. The \ac{rvpp} is located in southern Spain and includes a wind farm, two solar PV plants, and a flexible demand.
The production forecast data of the wind farm and the solar PV plants, representing a sample day of the spring season in Spain, are taken from~\cite{web:iberdrola_wind, BIENCINTO2022134821}. The solar PV plants and the wind farm each have a rated capacity of 50 MW and operating costs of 5 €/MWh and 10 €/MWh, respectively. A residential aggregator profile for the flexible demand is considered according to~\cite{oladimeji2022modeling}. The demand owner allows a 10\% tolerance for additional demand flexibility, which is allocated for the possible \ac{sr} provision. All energy forecast data related to \ac{rvpp} units is shown in Figure~\ref{fig:Energy_forecast}. The price forecast data for \ac{dam} and \ac{srm} are taken from the \ac{ree} website, and are shown in Figure~\ref{fig:Simulation_energy_reserve_price} for illustration purposes~\cite{web:ree_price}.

Two case studies are performed to analyze the performance of the proposed model. In the first case study, different values for all uncertain parameters related to the \ac{dam} and \ac{srm} electricity prices, \acp{ndrs} energy, and demand are considered to show the behaviour of the proposed model in different uncertain environments. In the second case, by means of an out-of-sample assessment, the bidding approach of this paper is compared with two models in the literature. The detailed description of the input parameters for the above cases is highlighted below:

\begin{itemize}
    \item Case 1.1: Deterministic case ($\Gamma ^{DA/SR}=\Gamma_{r}=\Gamma_{d}=0$);

    \item Case 1.2: Only the uncertainties of the energy of the \ac{ndrs} units and the demand are considered ($\Gamma ^{DA/SR}=0$ and $\Gamma_{r}=\Gamma_{d}=5$);
    
    \item Case 1.3: Only the \ac{dam} and \ac{srm} electricity price uncertainties are considered ($\Gamma ^{DA/SR}=5$ and $\Gamma_{r}=\Gamma_{d}=0$);    

    \item Case 1.4: Both \ac{dam} and \ac{srm} electricity price and energy uncertainties are considered ($\Gamma ^{DA/SR}=\Gamma_{r}=\Gamma_{d}=5$);

    \item Case 2: The results of the proposed model for $\Gamma ^{DA/SR}=\Gamma_{r}=0, 1, 2, ..., 9$ are compared with models in~\cite{Hadi} and~\cite{baringo2018day} using an out-of-sample assessment.
\end{itemize}

Simulations are performed on a Dell XPS with an i7-1165G7 2.8 GHz processor and 16 GB of RAM using the CPLEX solver in GAMS 39.1.1.

\vspace{-3mm}
\subsection{Case 1}
\label{sub:Case 1}

\begin{figure*}[!t]
    \centering
    
    \subfigure{\includegraphics[width=\textwidth]{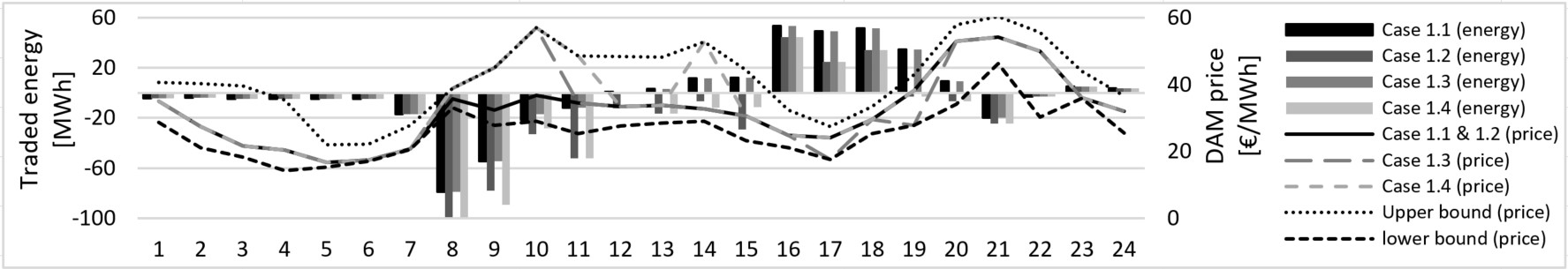}}
    \label{fig:Simulation_energy_price}

    \vspace{-5mm}
    
    \subfigure{\includegraphics[width=\textwidth]{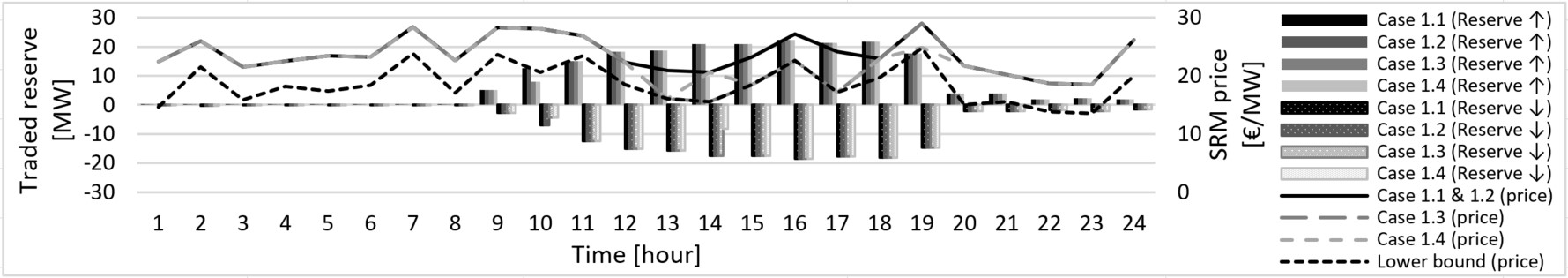}}
    \label{fig:Simulation_reserve_price}

    \vspace{-5mm}
    \vspace{-.5em}
    \caption{
    \ac{rvpp} traded energy and reserve versus electricity price in different case studies.}   
    \vspace{-1em}
     \vspace{-.8mm}    \label{fig:Simulation_energy_reserve_price}
\end{figure*}

Figure~\ref{fig:Simulation_energy_reserve_price} shows the \ac{rvpp} traded energy and reserve versus the electricity price for Cases 1.1 through 1.4. The general results for all cases show that between hours 8-11, when the demand is high and the production of \ac{ndrs} units is not enough to supply all demand, the \ac{rvpp} is an energy buyer in the electricity market. Between hours 12-15, although the demand is high, the production and demand of \ac{rvpp} are approximately equal, and \ac{rvpp} does not trade too much energy in most cases. However, in these hours the consideration of different uncertain parameters in Cases 1.1 through 1.4 has a significant effect on the trading direction of \ac{rvpp} (whether \ac{rvpp} is a seller or a buyer of energy). Between hours 16-19, as the demand decreases, the \ac{rvpp} becomes a seller of energy in most of the cases. The results for traded \ac{sr} shows that between hours 9-20, that \ac{rvpp} has high production, it provides more up and down \ac{sr} to the market.

The total sold energy of \ac{rvpp} in Cases 1.2 through 1.4 is decreased by 52.0\%, 0\%, and 51.7\%, respectively, compared to Case 1.1, whereas the total bought energy of \ac{rvpp} is increased by 74.2\%, 0\%, and 66.3\%, respectively. The total up (down) \ac{sr} provided by \ac{rvpp} in Cases 1.2 through 1.4 is decreased by 0 (0)\%, 2.2 (1.5)\%, and 7.5 (7.1)\%, respectively, compared to Case 1.1.

The results for each case study show that in the deterministic case (Case 1.1) and in the hours when \ac{rvpp} is an energy seller in the market, \ac{rvpp} usually sells more energy and reserve than in Cases 1.2 and 1.4. However, if \ac{rvpp} is an energy buyer, the energy bought in Case 1.1 is usually less than in Cases 1.2 and 1.4. The reason is that in the deterministic case, the \ac{rvpp} always takes an optimistic approach because it does not consider any uncertain parameter. In Case 1.2, considering the energy deviation of \acp{ndrs} and demand results in a lower amount of energy sold and a higher amount of electricity purchased from the market compared to Case 1.3, where only \ac{dam} and \ac{srm} electricity price uncertainties are considered. According to the comparison of Cases 1.1 and 1.3, considering only the electricity price uncertainty results in a lower amount of purchased energy only in some hours, e.g. hour 10, compared to Case 1.1. The reason is that this hour is one of the hours in which the electricity price goes to its worst case. Therefore, the \ac{rvpp} prefers to supply its demand with its production and also to provide less \ac{sr}. In other hours (except hours 1-6 and 12 with small differences) there are not too many differences between \ac{rvpp} traded energy in Cases 1.3 and 1.1. The reason is that although the electricity price goes to the worst cases in some hours in Case 1.3, the \ac{rvpp} must supply its demand or it can sell energy to the market with lower benefit. Considering all energy and price uncertainties in Case 1.4 results in a different bidding approach in some hours (e.g., hours 9, 10, 12, 14, and 15) compared to Case 1.2, which considers only energy uncertainty. Hours 9, 10, and 14 are exactly the hours where the worst cases of electricity price occur, forcing the \ac{rvpp} to increase or decrease its bid amount. Note that the worst cases of \ac{dam} electricity price occur in hours 8-11 and 14 (which are different from the worst cases of electricity price in Case 1.3).

\vspace{-3mm}
 \subsection{Case 2}
\label{sub:Case 2}
\begin{figure}
    \centering  \includegraphics[width=\columnwidth]{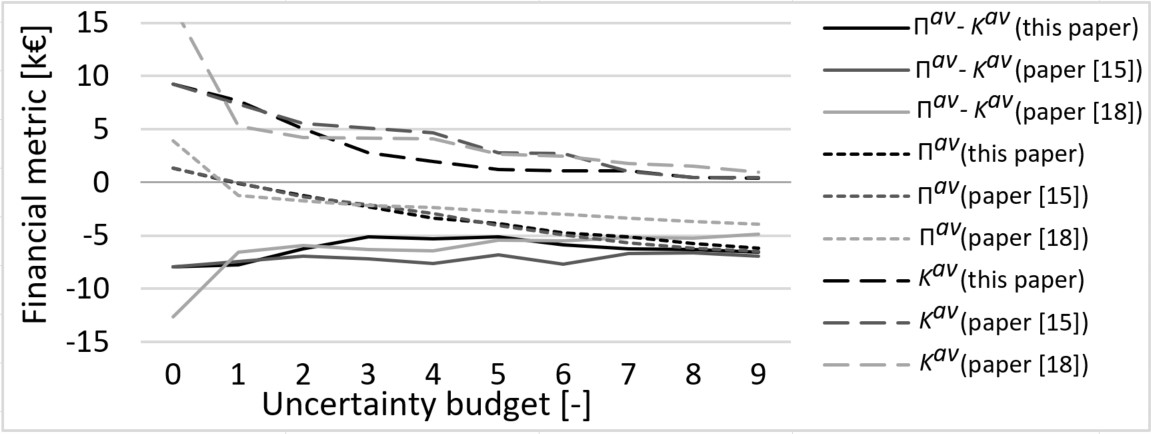}
    \vspace{-2em}
    \caption{The out-of-sample assessment for proposed model and the models in~\cite{Hadi} and~\cite{baringo2018day} ($\Gamma ^{DA/SR}=\Gamma_{r}=0, 1, 2, ..., 9$).}
    \label{fig:Out_of_Sample_fig}
    \vspace{-1em}
\end{figure}

Figure~\ref{fig:Out_of_Sample_fig} compares the results of an out-of-sample assessment for the proposed model and models in~\cite{Hadi} and~\cite{baringo2018day} for different values of uncertainty budgets between 0 and 9. In the figure, $\Pi^{av}$ represents the operating profit (no penalization applied), $K^{av}$ is the penalization cost for not complying with the energy bid, and the net profit of \ac{rvpp} is represented by $\Pi^{av}-K^{av}$. In~\cite{Hadi}, the energy robustness approach is adopted. The authors in~\cite{baringo2018day} use a multi-level optimization problem which implements an \ac{ro} approach to model the \ac{ndrs} units uncertainties, and a \ac{so} to capture the electricity price uncertainties. Therefore, the uncertainty budget in Figure~\ref{fig:Out_of_Sample_fig} for model~\cite{baringo2018day} refers only to the production of \ac{ndrs} units. To model the price uncertainty in their \ac{so} model, 200 scenarios are considered according to the \ac{ree} website~\cite{web:ree_price}. For the out-of-sample assessment, 1000 scenarios are generated based on the hourly distributions of uncertain parameters related to the \ac{dam} and \ac{srm} electricity prices and \acp{ndrs} production. The Weibull distribution, with its ability to model different degrees of skewness and tails, is used to generate scenarios to better capture the asymmetric behavior of uncertain parameters. Note that an equal value for all time periods, such as in~\cite{Hadi} and~\cite{baringo2018day}, can be considered for the penalty cost. However, the penalty cost related to the energy that is not provided is set to three times the \ac{dam} median price forecast in this paper. In this way, the deviation in the hours when the electricity price is higher leads to more penalty for \ac{rvpp}.
 
The net profit of \ac{rvpp} for uncertainty budget 0 in the proposed model and model~\cite{Hadi} is the same as the deterministic components in both models are the same. However, the model in~\cite{baringo2018day} results in a lower value of net profit for uncertainty budget 0 compared to the proposed model and~\cite{Hadi} due to the use of a different reserve provision strategy. In this paper and in~\cite{Hadi}, the production plus the reserve provided by each \ac{rvpp} unit is limited by the maximum production of each unit, while in~\cite{baringo2018day} this constraint is not defined and only the reserve provision limit by the entire \ac{rvpp} is considered. Therefore, for an uncertainty budget of 0, using the proposed model or the model in~\cite{Hadi} results in a lower energy bid in the \ac{dam} in several hours compared to~\cite{baringo2018day}. By increasing the uncertainty budget, the proposed model leads to a higher net profit obtained compared to model~\cite{Hadi}. 

The better results in terms of net profit by using the model~\cite{baringo2018day} compared to the proposed model is, to some extent, expected. This is due to the use of a more sophisticated approach to find the worst case of uncertainties of \acp{ndrs} production, and the consideration of the possibility of rescheduling the \ac{rvpp} units in the third level of the model~\cite{baringo2018day}. However, the proposed model shows a closely aligned results compared to~\cite{baringo2018day} even in some cases the obtained results in the proposed model are better than model in~\cite{baringo2018day} (see e.g. results for uncertainty budgets 0, 3, 4, and 5).

From the computational standpoint, the simulation time of different cases of the model~\cite{Hadi} is less than 2 s due to the simplified approach to identify the worst case of the optimization problem. The simulation time of the model proposed in this paper is less than 90 s in all cases, which meets the acceptable criteria for the \ac{rvpp} bidding problem when, e.g., different strategies need to be analyzed and compared before submitting the bid to the market operator. The simulation time of the model~\cite{baringo2018day} reaches 90 min in some cases. In summary, the proposed approach demonstrates outstanding computational performance against more intricate approaches such as~\cite{baringo2018day}, while providing results that compete in terms of profits with the model in~\cite{baringo2018day} and reducing the risk of penalization compared with~\cite{Hadi}.    

\section{Conclusion}
\label{sec:Conclusion}
In this paper, a novel, computationally efficient, single-level robust bidding method is proposed to capture multiple uncertainties in the \ac{dam} and \ac{srm} electricity prices as well as \ac{ndrs} production and demand of an \ac{rvpp}. The non-linear couplings between different uncertainties in the objective function and constraints of the optimization problem are addressed by developing an accurate linear model based on the big M method. The obtained results show that the uncertainty of \ac{ndrs} and demand has the highest impact on the bidding approach of \ac{rvpp} compared to the electricity price uncertainty. Furthermore, the sensitivity analysis shows that the \ac{rvpp} operator can significantly increase its net profit by considering even a low or median value for the risk measure parameter (uncertainty budget). In addition, the simulation results show the computational efficiency of the proposed model as well as high consistency results with more complicated multi-level models. In the future works, the authors aim to optimize the risk measure parameter so that the \ac{rvpp} operator obtains a certain desired profit.

\vspace{-2mm}

\bibliographystyle{IEEEtran}
\bibliography{refs.bib}

\clearpage

\end{document}